\documentclass[conference,letterpaper]{IEEEtran}

\usepackage{url}
\usepackage{dsfont}
\usepackage{tikz}
\usetikzlibrary{quantikz2}
\usepackage{tikzsymbols}
\usepackage[most]{tcolorbox}
\usepackage{booktabs}
\usepackage{enumitem}

\definecolor{bluebsc}{RGB}{31,57,112}

\begin{document}

\title{Distributed Quantum Circuit Cutting for Hybrid Quantum-Classical High-Performance Computing}

\author{
    \IEEEauthorblockN{Mar Tejedor\IEEEauthorrefmark{1}, Berta Casas\IEEEauthorrefmark{1}\IEEEauthorrefmark{2}, Javier Conejero\IEEEauthorrefmark{1}, Alba Cervera-Lierta\IEEEauthorrefmark{1}, Rosa M. Badia\IEEEauthorrefmark{1}\IEEEauthorrefmark{3}}\\
    \IEEEauthorrefmark{1}Barcelona Supercomputing Center, Barcelona, Spain\\
    \IEEEauthorrefmark{2}Universitat de Barcelona, Barcelona, Spain\\
    \IEEEauthorrefmark{3}Universitat Politècnica de Catalunya, Barcelona, Spain\\
    mar.tejedor@bsc.es\\
}

\maketitle

\begin{abstract}
Most quantum computers today are constrained by hardware limitations, particularly the number of available qubits, causing significant challenges for executing large-scale quantum algorithms. Circuit cutting has emerged as a key technique to overcome these limitations by decomposing large quantum circuits into smaller subcircuits that can be executed independently and later reconstructed. In this work, we introduce Qdislib, a distributed and flexible library for quantum circuit cutting, designed to seamlessly integrate with hybrid quantum-classical high-performance computing (HPC) systems. Qdislib employs a graph-based representation of quantum circuits to enable efficient partitioning, manipulation and execution, supporting both wire cutting and gate cutting techniques. The library is compatible with multiple quantum computing libraries, 
including Qiskit and Qibo, and leverages distributed computing frameworks to execute subcircuits across CPUs, GPUs, and quantum processing units (QPUs) in a fully parallelized manner. We present a proof of concept demonstrating how Qdislib enables the distributed execution of quantum circuits across heterogeneous computing resources, showcasing its potential for scalable quantum-classical workflows.
\end{abstract}


\section{Introduction}
\label{sec:introduction}
In recent years, quantum computing has become an important technology with the potential to revolutionize a wide range of fields. With the rapid advancements in quantum hardware and algorithms, the ability to efficiently execute large-scale quantum computations is essential for achieving the full potential of quantum computing in practical applications. However, as quantum algorithms grow in size and complexity, the resources required to execute them often exceed the capabilities of current quantum devices. State-of-the-art quantum computers are not fault-tolerant, meaning they have a limited number of qubits with short coherence times and uncorrected errors. Therefore, two major challenges arise in the design of quantum algorithms for near-term devices: how to mitigate quantum errors and how to scale solutions to problems involving large numbers of qubits on small devices.

Quantum computers are not designed to replace traditional computation. In fact, all quantum algorithms require pre- and post-processing steps that depend on traditional (classical) computing. Moreover, there exists a vast family of near-term quantum algorithms that include a quantum-classical optimization loop to converge towards a solution~\cite{NISQ}. Therefore, quantum computers are envisioned as another partition within a large cluster of classical computers, fully integrated into a supercomputing infrastructure, where they will tackle some parts of large computational problems.

The technology maturity of quantum computing is far behind traditional supercomputing. Various technological platforms exist for building quantum computers, each of them with different operational needs and performance metrics. Recently, quantum computers have started to be integrated into high-performance computing (HPC) infrastructures, presenting several challenges. One key challenge is managing the response times and availability of quantum devices while executing large-scale hybrid algorithms. Even with current advancements, it seems unlikely that all qubits will fit on a single quantum chip. In fact, quantum communication protocols are being developed alongside quantum computing to connect quantum computers through the so-called \textit{quantum internet} \cite{Wei2022}. Both for current quantum-HPC integration needs and for long-term quantum computing connections, one will need to design an optimized scheduler capable of controlling and operating all nodes - both classical and quantum - within a supercomputer environment. 

In this work, we aim to address the challenge of orchestrating hybrid quantum-classical applications through \texttt{Qdislib}, an open-source distributed and flexible High Performance Quantum Library. Based on the task-based programming model PyCOMPSs~\cite{badia2015comp}, \texttt{Qdislib} efficiently manages the execution both in quantum and classical systems.
The first features provided in \texttt{Qdislib} and presented in this paper are quantum circuit cutting techniques that split large quantum circuits into smaller ones, either to fit current term quantum computers or to allow classical simulation of their results in a distributed environment. 

The paper is organized as follows.  First, we present the basic concepts to understand the contributions of this work, namely the quantum computing limitations to be simulated classically and distributed, the quantum circuit cutting techniques and the PyCOMPSs programming model.
Following this, the \texttt{Qdislib} overview section provides a detailed examination of \texttt{Qdislib}, a large scale High-Performance Quantum library, including its design and implementation. In the subsequent Evaluation section, we present the findings and benchmarks conducted to assess \texttt{Qdislib}'s performance across various quantum computing tasks, comparing it with circuit simulation on HPC devices. 
Next, we review some related works that also tackle quantum circuit cutting techniques from a HPC perspective.
Finally, the Conclusion section offers a summary of our contributions and suggests potential avenues for further investigation. 

\begin{tcolorbox}[
 colback=blue!5!white,colframe=bluebsc,
 colbacktitle=bluebsc,
 title=The main contributions of this work are:]
\begin{itemize}[leftmargin=*]
\item The extension of distributed task-based environment to simultaneously control simulations in production CPU and GPU systems, as well as on real quantum systems.
\item The proposal of novel techniques for finding partitions of large quantum circuits, which adapt to state-of-the-art technological needs and address the challenge of circuit scalability in quantum computing. 
\item A demonstration of the performance of \texttt{Qdislib} for representative quantum circuit-cutting applications, employing both CPU and GPU environments, as well as a digital quantum computer integrated into a supercomputer and a cloud-accessed quantum computer. In doing so, we contributed to and validated the full integration of a hybrid quantum and high-performance computing infrastructure.
    
\end{itemize}
\end{tcolorbox}

\section{Motivation and background}
\label{sec:background}
\subsection{Quantum computing fundamentals}
Quantum computing can be categorized into digital and analog approaches, with gate-based quantum computing being the most widely used and closest counterpart to classical computation. This work focuses on the digital paradigm, leveraging quantum circuits to encode and manipulate quantum information.

A quantum algorithm consists of a set of instructions involving the execution of one or more quantum circuits, which create and manipulate information encoded in quantum states. The fundamental units of quantun information, called \textit{qubits}, are the building blocks that store the quantum information. Mathematically, a single qubit is represented by a unit vector $|\psi\rangle$ in a two-dimensional complex Hibert space. To represent $n$ qubits, the joint quantum state is described by a unit vector in a $2^n$-dimensional Hilbert space. To evolve or manipulate quantum states, we apply a sequence of quantum gates, which are reversible physical operations represented mathematically by unitary matrices \cite{nielsen2000quantum}.

Simulating such quantum systems with traditional computation requires storing and manipulating large vectors and matrices scaling exponentially with the number of qubits $n$. This scaling quickly becomes intractable, reaching a limit where classical computers can no longer efficiently simulate quantum computations- a regime often referred to as \textit{quantum advantage}. Nevertheless, this motivates the development of more sophisticated simulation techniques \cite{Ors2014} and software to manage resources more efficiently and push the quantum advantage barrier further.

Quantum computing power lies in its ability to exploit the superposition of an exponential number of bit strings (all possible combinations of the internal states of qubits) and the interference created in this superposition when quantum gates are applied. The final step in a quantum circuit is the \textit{measurement}: the quantum state (possibly in a superposition) collapses to a single classical bit string. By executing and measuring the same circuit multiple times, one can estimate the probability distribution over bit strings. This distribution provides partial access to the information encoded in the output state of the circuit and it is used to compute expectation values of observables. Formally, for a given Hermitian operator $\hat E$, the expectation value is denoted by $\langle \hat E\rangle$.

Another fundamental property in quantum computing is entanglement, a type of correlation between qubits that does not have a classical counterpart. In quantum circuits, entanglement can appear through the application of quantum gates that act on multiple qubits. It prevents the independent description or manipulation of the quantum information stored in one subset of qubits without affecting the information contained in others. As a result, a quantum circuit cannot be arbitrarily divided into separate parts without disrupting the entangled quantum state. However, there exist techniques to perform such a partitioning (or circuit cutting), but they come with a computational overhead, as will be discussed in more detail in the following subsections.  

\subsection{Quantum circuit cutting}
Current-term quantum computers have a limited number of qubits per quantum processing unit (QPU) constrained by the technological developments of the physical platform. To scale quantum computational capabilities, most platforms will likely rely on interconnecting multiple QPUs through quantum or classical channels. Additionally, as quantum circuits increase in complexity (measured by the number of qubits and gates), scalability becomes a critical challenge. Circuit cutting has emerged as a promising technique for partitioning large quantum circuits into smaller, more manageable subcircuits, preserving computational coherence while partially mitigating scalability limitations. 
In this work, we focus on two methods of circuit cutting -- wire cutting~\cite{peng2020simulating,tang2021cutqc} and gate cutting~\cite{mitarai2021constructing} -- and assume classical communication between the resulting subcircuits.   

\subsubsection{\bf{Wire Cutting}} 
\begin{figure}[htbp]
\centerline{\includegraphics[width=0.4\textwidth]{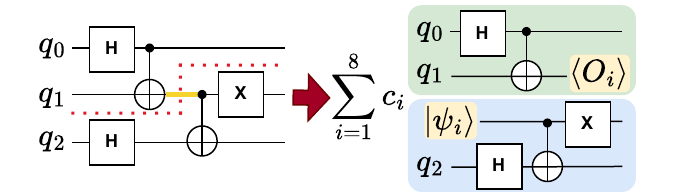}}
\caption{
Example of a quantum circuit partitioned. 
}
\label{fig:wire_cut}
\end{figure}

This approach consists of partitioning a quantum circuit into independent subcircuits by `cutting' a qubit wire at a chosen location. After the cut, we obtain a collection of subcircuits with different measurements and state-preparations, which are run separately. The final result is then reconstructed by appropriately combining the outputs~\cite{peng2020simulating, tang2021cutqc}.
As an introductory example, consider a single qubit circuit with two gates $U_1$, $U_2$ with a wire cut between them 
\begin{equation}
\begin{quantikz}
\lstick{$| 0 \rangle$} & \gate{U_1}
& \qw \slice{} 
& \qw & \gate{U_2} & \qw \rstick{$\langle \hat E\rangle$,} 
\end{quantikz}
\end{equation}
where the circuit's final output corresponds to the expectation value of an observable $\hat E$. By introducing the wire cut, we decompose the original circuit into two independent subcircuits. The first subcircuit applies $U_1$ and measures a given expectation value. The second subcircuit prepares an initial state before applying $U_2$ and then measures $\hat E$. The final expectation value of the entire circuit can be reconstructed by combining the outputs of these subcircuits with appropriate coefficients $c_i$, as illustrated in Figure~\ref{fig:wire_cut}. Using Figure \ref{fig:wire_cut} as an example, by cutting the second qubit wire the circuit is divided into two disconnected parts. For the top subcircuit, we need to compute expectation values of Pauli observables $O_{i}\in \{X,Y,Z,I\}$ \cite{nielsen2000quantum} replacing the wire, and in the bottom one, we initialize the cut wire with the eigenstates of these Pauli matrices $|\psi_{i}\rangle$. The coefficients $c_i\in\{+\frac 12, -\frac 12\}$ are chosen to ensure the correct reconstruction of the observable, as described in \cite{peng2020simulating}. In total, we require eight pairs of subcircuits for simulating the expectation value, i.e. the resources to reconstruct the output expectation value scale exponentially with the number of $k$ cuts as $8^k$.

\subsubsection{\bf{Gate Cutting}}
This circuit cutting technique enables the decomposition of entangling gates, that is, gates that can not be expressed as a tensor product of local unitaries, $U\neq U_1 \otimes U_2$. A general entangling gate $U$ can be decomposed as a linear combination of local operations: $U = \sum_{j = 1}^\chi c_j U_j^{(1)}\otimes U_j^{(2)}$, where $U^{(1)}_j$ and $U^{(2)}_j$ are local unitary gates and $c_j\in \mathds{C}$. The number of terms in the sum $\chi$ is related to the entangling power of the gate $U$. However, estimating the expectation values of an observable $\hat{E}$ in this decomposition requires evaluating off-diagonal terms of the form 
$\langle 0|(U_k^{(1)}\otimes U_k^{(2)} )^\dagger \hat E (U_j^{(1)}\otimes U_j^{(2)})|0\rangle$, which is experimentally challenging to implement when $k\neq j$. 
To overcome this limitation, in this work we adopt the quasi-probabilistic gate decomposition framework introduced in \cite{mitarai2021constructing}. With this approach, the gate is rewritten as a weighted sum over local operations with intermediate projective measurements. 
This method avoids the need to evaluate off-diagonal terms, but it may require hardware capabilities that support mid-circuit measurements and qubit reset, which may not be available in some quantum platforms. 

\begin{figure}[t]
    \centering
    \includegraphics[width=0.8\linewidth]{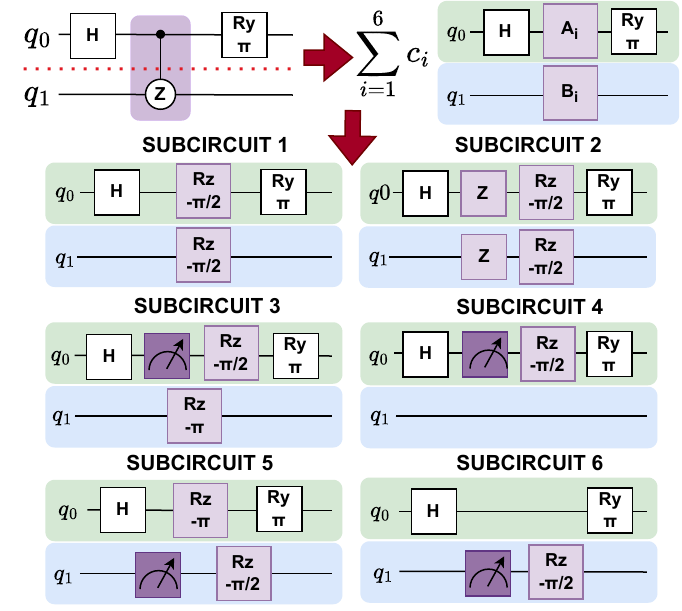}
    \caption{Decomposition for cutting a Controlled-Z gate. Six subcircuits are required to reconstruct this gate \cite{mitarai2021constructing}. 
    }
    
    \label{fig:gate_cut}
\end{figure}

As a concrete example, Figure~\ref{fig:gate_cut} illustrates the decomposition of a Controlled-Z (CZ) gate into six sub-circuits. Only four types of single-qubit gates are required (Haddamard $H$, Pauli-$Z$, $R_{z}(\theta)=e^{i\frac\theta2 Z}$, and $R_{y}(\theta)=e^{i\frac\theta2 Y}$).
The reconstruction of the expectation value relies on classical post-processing using coefficients $c_i$, which depend on the outcomes of the intermediate projective measurements (represented with the \begin{quantikz} \meter{}
\end{quantikz} symbol). 
This technique has also exponential scaling in the number of gate-cuts. For example, for the CZ gate, the number of circuits scales to $6^k$, where $k$ is the number of gate cuts. 

Since other commonly used two-qubit gates—such as the CNOT— can be obtained from the CZ gate via single-qubit unitaries, this decomposition can be naturally extended to those gates as well. For a detailed theoretical derivation of this decomposition for general two-qubit gates see \cite{mitarai2021constructing}.

\subsection{PyCOMPSs}

\begin{figure*}[t]
    \centering
    \includegraphics[width=0.85\linewidth]{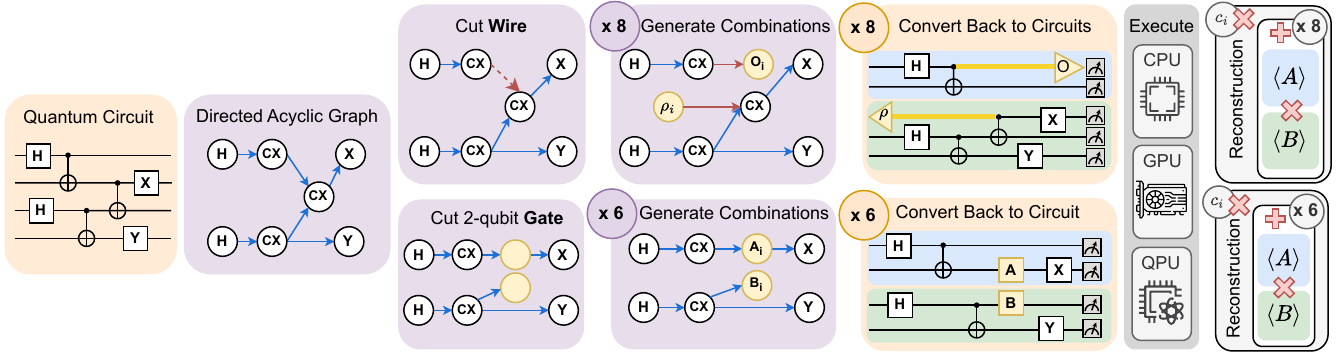}
    \caption{Circuit Cutting \texttt{Qdislib} workflow.}
    \label{fig:workflow_qdislib}
\end{figure*}

PyCOMPSs~\cite{tejedor2017pycompss,badia2015comp} is a task-based programming model and runtime designed for parallel execution of workflows across diverse computing resources, including CPUs, GPUs, and distributed systems. It enables dynamic scheduling of tasks, allowing the efficient distribution of workloads based on the available resources. PyCOMPSs simplifies the development of parallel applications by automatically handling data dependencies and distributing tasks without the need for explicit resource management. It has been used in multiple application areas such as biomedicine, engineering, biodiversity, chemistry, astrophysics, finance, telecommunications, manufacturing, natural hazards and earth sciences~\cite{castro2020hunting,
bonas2018re,
andrio2019bioexcel,
lezzi2012enabling,
ejarque2024managing,
monterrubio2024machine,
elia2023end}. 

In the context of quantum computing, PyCOMPSs has been previously used to parallelize the RosneT, a library for distributed, out-of-core block tensor algebra~\cite{sanchez2021rosnet}. 
PyCOMPSs is particularly useful in  hybrid quantum-classical computing for dynamically scheduling tasks  between classical and quantum resources. This flexibility allows for optimal use of computational resources, particularly in scenarios where different parts of a quantum circuits or algorithms can benefit from being executed on different types of hardware.

In our work, PyCOMPSs will be used for the manipulation and dynamic allocation of sub-circuits in the circuit cutting process, ensuring that each segment is executed in parallel and on the most suitable resource—whether it be a CPU, GPU, or quantum processor—thereby optimizing performance and reducing overhead.

\section{\texttt{Qdislib} overview}
\label{sec:qdislib overview}
\texttt{Qdislib} is a flexible and efficient library designed for quantum circuit cutting, enabling the execution of large quantum circuits by partitioning them into smaller subcircuits. This approach allows circuits to be simulated or executed on hardware with limited qubit resources, improving scalability and compatibility with various quantum computing platforms. By exploiting a graph-based representation of quantum circuits, \texttt{Qdislib} provides a robust 
method for identifying cuts and partitioning circuits. The library has been designed to integrate seamlessly with multiple quantum programming languages, starting with Qiskit \cite{qiskit2024} and Qibo~\cite{qibo_paper}, and supports execution on both quantum hardware (either cluster or cloud) and classical HPC systems.

\subsection{Graph-Based Approach for Circuits}
\texttt{Qdislib} implements circuit cutting using a graph-based approach, where quantum circuits are represented as directed acyclic graphs (DAGs). This abstraction offers a flexible, software-agnostic framework for circuit manipulation and facilitates compatibility with various quantum computing libraries. When a circuit is submitted, \texttt{Qdislib} first converts it into a DAG—each node corresponds to a quantum gate, and edges represent the gate execution order. All subsequent operations, such as wire or gate cutting, are performed directly on this graph representation, enabling efficient modification and identification of parallelizable segments.

Once the necessary transformations are applied, the graph is converted back into a quantum circuit in the appropriate language, depending on the execution context (simulation or real hardware). If the original language lacks features like intermediate measurement—essential for gate cutting—\texttt{Qdislib} automatically translates the circuit into a supported language to ensure correct execution. This abstraction ensures consistent implementation of quantum operations regardless of the input language and extends to diverse execution backends, from classical HPC-based simulations to various quantum hardware platforms.

\subsection{Wire and Gate Cutting in \texttt{Qdislib}}
\texttt{Qdislib} implements a comprehensive circuit cutting methodology that handles both wire cutting and gate cutting through an optimized workflow (Figure \ref{fig:workflow_qdislib}). As said before, the process begins by converting the input quantum circuit into a DAG. After that, users can then specify cutting points either manually or through the \textit{FindCut} function provided by the library and described in the following section. For wire cutting, this involves defining tuples of gates between which connections should be cut, while gate cutting requires selecting a list of specific two-qubit gates.

In the graph representation, the system generates all valid subcircuit combinations, according to the cut type, with wire cutting producing $8^k$ combinations and gate cutting producing $6^k$ combinations, where $k$ represents the number of cuts (see Section~\ref{sec:background} for details). This exponential scaling reflects the expected fundamental trade-off between circuit decomposition and computational overhead.

A critical distinction between the two approaches lies in their decomposition requirements since gate cutting introduces intermediate measurements as part of the gate decomposition process (see Fig.~\ref{fig:gate_cut}). \texttt{Qdislib} automatically handles these technical requirements by adapting the quantum library used and the resources where to execute it,  whether a simulator or physical quantum hardware.

After generating all the needed graph combinations, each one is converted back into executable quantum circuits. The library leverages PyCOMPSs framework through all the workflow to support parallel execution across multiple cores and nodes. This enables the concurrent generation and execution of multiple subcircuits,  significantly reducing the total computation time.

The final stage consists of reconstructing the expected value of the original circuit by combining the measurement results of all subcircuits. This employs reconstruction formulas based on quasi-probability decomposition~\cite{mitarai2021constructing} and wire cutting reconstruction~\cite{peng2020simulating}. The framework provides configurable precision parameters that allow users to balance statistical reconstruction accuracy against computational cost adapting the number of shots in the simulation.
This process ensures that the results of the cut procedure are equivalent to the behavior of the original circuit but with the computation split into smaller, more manageable subcircuits.

\subsection{\textit{FindCut} Algorithm in \texttt{Qdislib}}
The \texttt{Qdislib} library introduces the \textit{FindCut} algorithm, a tool designed to automatically search the best cut of a quantum circuit given specific user-defined constraints. Its goal is to partition circuits in a way that enables parallel execution, minimizes the number of cuts and qubits per subcircuit, and controls subcircuit complexity. The algorithm accepts a quantum circuit along with parameters that define specific cutting constraints—such as the maximum number of qubits per subcircuit or the allowed number of cuts. Based on these inputs, it searches for the most efficient partitioning that satisfies the given requirements, aiming to reduce resource consumption and improve execution efficiency.

\begin{figure}[htbp]
    \centering
    \includegraphics[width=1\linewidth]{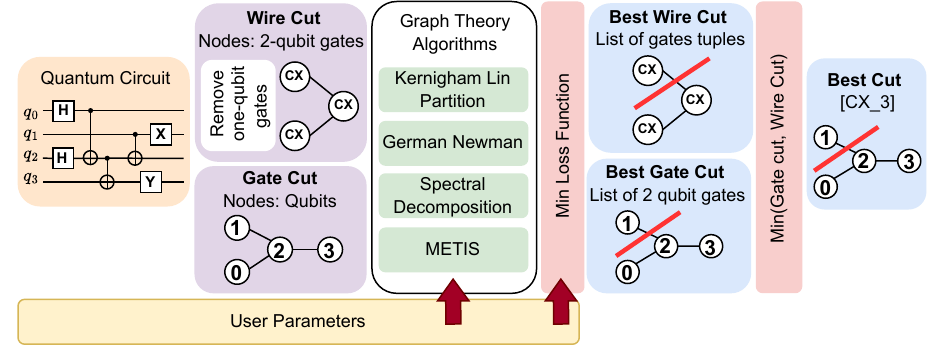}
    \caption{\textit{FindCut} workflow in \texttt{Qdislib}.} 
    \label{fig:FindCut}
\end{figure}

As shown in Figure \ref{fig:FindCut}, and as previously discussed, the algorithm begins by transforming the quantum circuit into a graph representation. For gate cutting, nodes represent qubits and edges correspond to two-qubit gates. In contrast, wire cutting starts by removing all single-qubit gates; the graph is then defined with nodes as two-qubit gates and edges indicating their connectivity in the circuit. This distinction in representation ensures that the algorithm correctly models the dependencies and interactions within the circuit for each cutting method.

To search for best circuit cuts, the \textit{FindCut} algorithm employs several graph partitioning techniques, executed in parallel for both gate and wire cutting. Specifically, it integrates the Kernighan-Lin \cite{kernighan1970efficient}, Girvan-Newman \cite{newman2004finding}, spectral decomposition \cite{chung1997spectral}, and METIS \cite{karypis1998fast} algorithms. The first three methods are implemented via the \texttt{NetworkX} library \cite{networkx}, while METIS is accessed through its dedicated library. Each technique contributes a different strategy for effective partitioning: Kernighan-Lin iteratively swaps node pairs to minimize edge cuts; Girvan-Newman removes edges with the highest betweenness centrality to reveal strongly connected subcircuits; spectral decomposition leverages the Laplacian eigenvalues of the graph for balanced clustering; and METIS applies a multilevel approach that coarsens, partitions, and refines the graph for efficient cutting. These implementations are integrated into the \textit{FindCut} function in \texttt{Qdislib}, allowing for flexible and constraint-aware circuit partitioning.

Once all partitions are computed, the algorithm evaluates them using a loss function designed to prioritize circuit cuts that minimize resource overhead while enhancing computational efficiency. The loss function is defined as:
\begin{equation}
    \text{Loss} = \alpha \cdot \text{min\_cuts} + \beta \cdot \text{max\_comp} + \gamma \cdot \text{min\_qubits}.
\end{equation}

\begin{itemize}
\item \text{min\textunderscore cuts}: Minimizes the number of gates and wires cut.
\item \text{max\textunderscore comp}: Maximizes the number of independent components (subcircuits) formed by the cut.
\item \text{min\textunderscore qubits}: Minimizes the number of qubits in each subcircuit.
\end{itemize}
The coefficients $\alpha$, $\beta$, and $\gamma$ are weights that control the importance of each term in the optimization process. By default, these are set to values that produce a well-balanced cut, but they are fully configurable to suit different use cases or hardware constraints. After computing the best cuts from both gate and wire cutting strategies, the algorithm compares their respective loss scores and selects the cut with the lowest loss. This ensures the chosen partitioning strategy strikes an effective balance between circuit decomposition and execution efficiency.

Additionally, the \textit{FindCut} algorithm provides several key parameters that allow the user to control how the circuit is partitioned:
\begin{itemize}[leftmargin=*]
    \item \textbf{max\_qubits}: Specifies the maximum number of qubits allowed in each subcircuit. For example, if the user’s quantum hardware supports a maximum of 5 qubits, the algorithm ensures that no subcircuit exceeds this limit.
    \item \textbf{max\_components}: Limits the number of independent subcircuits that can be created. If splitting a circuit into too many subcircuits would increase execution overhead, this parameter allows the user to set a maximum number of components.
    \item \textbf{max\_cuts}: Controls the maximum number of cuts allowed by the algorithm. Since each additional cut exponentially increases the number of subcircuits, this parameter lets the user balance computational cost and performance.
    \item \textbf{wire\_cut} and \textbf{gate\_cut}: These boolean parameters enable the user to specify whether the algorithm should consider wire cuts, gate cuts, or both. The algorithm then searches for the best cut type based on the user’s preferences and returns the best possible solution.
\end{itemize}

\begin{tcolorbox}[colback=blue!5!white,
colframe=bluebsc,
colbacktitle=bluebsc,
title={\textit FindCut} advantages:]
The \textit{FindCut} algorithm in \texttt{Qdislib} offers key advantages for efficient quantum circuit partitioning. It provides flexibility by allowing users to define constraints suited to their quantum hardware or use case. 
The highly parallelized algorithm explores multiple cuts quickly, while the loss function optimizes cuts to balance minimal qubit usage and maximum parallelism,  making it a powerful tool for optimizing quantum circuits for hybrid quantum-classical execution environments.
\end{tcolorbox}

\section{Evaluation}
\label{sec:evaluation}
In this section, we outline the benchmarking methodology employed to evaluate the performance of the \texttt{Qdislib} in managing large-scale quantum computations. 
We selected two representative benchmarks that comprise many circuits' features
and carefully and thoughtfully assess the efficiency of \texttt{Qdislib} in addressing the challenge of circuit scalability through circuit-cutting techniques. 
Although \texttt{Qdislib} supports both wire and gate cutting, we test the performance of using gate cuts only, as we identify that most quantum computing applications will require significantly more cuts when using wire cutting than if using gate cutting. This is because most applications have certain qubit connection structures that make gate cutting more efficient than wire cutting.

First, we select the hardware-efficient ansatz (HEA) quantum circuit. This broadly used circuit balances current quantum hardware constraints with circuit complexity to address several types of applications \cite{NISQ}. These types of circuits present a certain qubit connectivity structure which helps us to envision the necessary computational resources in other specific benchmarks such as the Quantum Fourier Transform, graph states, or the generation of specific quantum states.
Second, we select the random quantum circuits from the Google's quantum supremacy work \cite{arute2019quantum} to test \texttt{Qdislib}'s performance in handling highly complex quantum circuits that are hard to simulate classically. Quantum applications, such as simulating condensed matter systems, may involve similar complexity in qubit connectivity and circuit depth as these random circuits. In both benchmarks, the angles of the parameterized quantum gates are randomly initialized, and we compute the expectation value of the $Z^{\otimes n}$ operator, where $n$ is the number of qubits in the full quantum circuit. 

\begin{tcolorbox}[colback=blue!5!white,
colframe=bluebsc,colbacktitle=bluebsc,
title={Computational backends:}]
Our evaluation considers four types of computational backends, with the possibility of combining all them in a single application execution. On the classical computing side, we use general-purpose processors ({\bf CPUs}) and accelerated processors ({\bf GPUs}). On the quantum side, we utilize digital superconducting circuit-based quantum processors ({\bf QPUs}), accessed both 
within the {\bf same infrastructure} as the classical devices and via {\bf cloud services}.
\end{tcolorbox}

We evaluate the performance of \texttt{Qdislib} in cutting quantum circuits of varying sizes, simulating them in a distributed environment across multiple nodes, and executing them on real quantum hardware. Additionally, we explore hybrid execution strategies, where some subcircuits are simulated classically while others are executed on real QPUs.

To simulate quantum circuit executions on classical hardware, we use the default Qiskit-Aer simulator \cite{qiskit2024}, which operates on the full wavefunction and supports CPU-based execution. For GPU-based simulation, we employ the cuStateVec simulator from NVIDIA’s cuQuantum framework \cite{Bayraktar2023}. For all executions, we set the number of shots to 1024 to ensure consistent conditions across simulations.

\subsection{Execution infrastructure}
The experiments have been performed in \textit{MareNostrum 5}, a pre-exascale EuroHPC supercomputer that includes multiple partitions. The general purpose partition (MN5-GPP) has 6,408 nodes with 2 Intel Xeon Platinum 8480, with a total of 112 cores at 2GHz. The accelerated partition (MN5-ACC) includes 1120 nodes with two Intel Xeon Platinum 8460Y+ each (80 cores per node) and  4 NVIDIA Hopper H100 64GB HBM2 per node. 

Another partition of the system is a digital gate-based quantum computer based on superconducting circuit technology. This quantum computer is part of the \textit{MareNostrum Ona}.
At the time this work was written, it operated a 5-qubit quantum chip, with plans to upgrade to chips supporting up to 35 qubits. The qubits are arranged in a star topology, with a central qubit connected to the other four.  
A login node connected to all the control electronics of the quantum computer supports its operation. The current configuration is based on Slurm to submit quantum circuit descriptions for execution. While the GPP and ACC partition share a single Slurm instance, Ona is served by a separated one, although a slurm federation instance is planned in the longer term. 
A shared file system based on GPFS connects all \textit{MareNostrum 5} partitions: GPP, ACC and Ona.  

The software libraries used to program this quantum computer are Qibo \cite{qibo_paper} (open-source, for gate-level programming) and 
Qililab \cite{qililab} (for pulse-level programming, characterization and calibration). 
Qibo programs automatically translate to Qililab programs which, in turn, are compatible with the middleware, i.e. the software that operates the controlled electronics that transform the abstract quantum instructions into the quantum hardware instructions. 

IBM offers a wide variety of quantum computers through their IBM Quantum cloud access. In particular, for this work, we use the   \textit{IBMQ-Marrakesh} device composed of 156 superconducting qubits. The qubits are arranged in rectangles of 3 qubits tall and 6 qubits wide, connected among them forming a heavy hexagonal topology (for details, check the IBMQ documentation~\cite{IBMQ}).
The quantum library used to operate these quantum computers is Qiskit \cite{qiskit2024}.

\subsection{Execution environment}
The execution environment consists of distributed heterogeneous resources, including CPUs, GPUs, and QPUs, with different computing architectures and different service models (clusters and cloud). BSC hosts the \textit{MareNostrum 5} supercomputer with the GPP, ACC and ONA partitions, and IBM provides the IBM Quantum Cloud. These diverse resources are leveraged cooperatively by PyCOMPSs through  the circuit-cutting algorithms implemented in the \texttt{Qdislib}. 
The main challenge for the classical-quantum execution has been overcoming communication barriers between infrastructures.

\begin{figure}[htbp]
    \centering
    \includegraphics[width=1\linewidth]{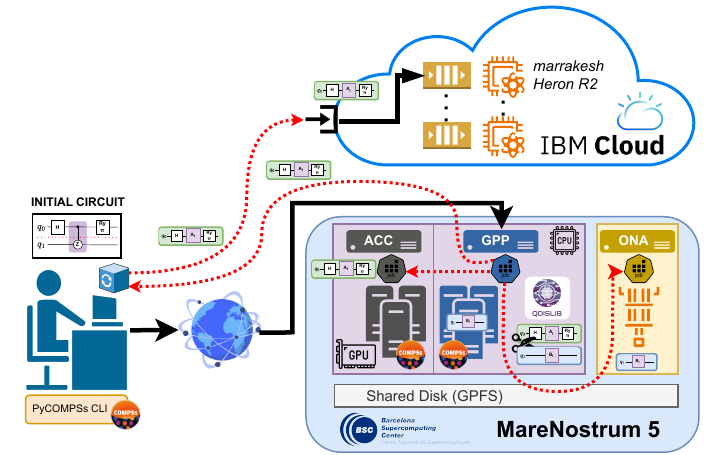}
    \caption{Hybrid execution schema.}
    \label{fig:hybrid_schema}
    
\end{figure}

Figure \ref{fig:hybrid_schema} illustrates how these resources are jointly utilized during the execution.
The process starts by submitting a quantum circuit for execution with \texttt{Qdislib}. This is is submitted to \textit{MareNostrum 5} via the PyCOMPSs CLI, a PyCOMPSs user interface that supports the remote execution of PyCOMPSs applications. This triggers a single Slurm job within \textit{MareNostrum 5} (ACC or GPP partition) that performs the circuit cutting, splitting the circuit into smaller subcircuits. The library is designed to handle tasks differently depending on the underlying resource type (CPU, GPU, QPU Ona, and QPU Cloud). This allows the PyCOMPSs runtime to efficiently orchestrate the execution across these different resources. For instance, since GPP and ACC are partitions within the same cluster, a single job can utilize resources from both partitions, such as CPUs from GPP and ACC, and/or GPUs from ACC. However, the Ona system, which operates on its own independent job scheduler, requires an agent that continuously listens for circuit execution requests intended for the Ona QPU. This agent is designed to run indefinitely, waiting for incoming requests. Communication with the agent is facilitated through the GPFS shared disk, where files are used to transfer data between the agent and other components of the system.

The most complex integration challenge arises with IBM Quantum Cloud, due to the connectivity limitations between \textit{MareNostrum 5} and remote cloud resources. To address this, QPU Cloud tasks are first sent to the CPUs of \textit{MareNostrum 5}, which then interact with the IBM Cloud API through a double port forwarding mechanism before being executed on the QPU Cloud. This process routes traffic from a \textit{MareNostrum 5} worker node to the login node and ultimately to the user’s laptop. On the laptop, a proxy server is configured to forward requests between \textit{MareNostrum 5} and the IBM Cloud, ensuring smooth and bidirectional communication.

These engineering decisions have allowed the seamless and successful utilization of all available resources for hybrid quantum-classical executions.

\subsection{Benchmark 1: Hardware Efficient Ansatz}
The first benchmark consists of cutting a quantum circuit with a Hardware-Efficient structure (\textit{Hardware-Efficient ansatz} (HEA)). The quantum computing community employs this particular circuit architecture to design parameterized quantum circuits leveraging state-of-the-art hardware capabilities. There are many proposed HEA designs, all of them sharing a layered structure: each layer is composed of single-qubit gates applied to all qubits in parallel, followed by two-qubit gates that entangle the qubits in a ladder-like pattern. This layer is applied $L$ times, increasing the quantum circuit's depth and complexity. In particular, we use the HEA obtained from the \textit{he\_circuit} function from Qibo \cite{qibo_paper}
that employs single-qubit rotational gates and CZ two-qubit gates. The total circuit depth scales linearly with the number of qubits $n$ and the number of layers $L$, following the expression $(2+n)L$. When the circuit is partitioned into multiple subcircuits, each subcircuit retains the same linear depth scaling, $(2+m)L$, where $m$ is the number of qubits of the subcircuit.

\begin{figure}[htbp]
    \centering
    \includegraphics[width=0.7\linewidth]{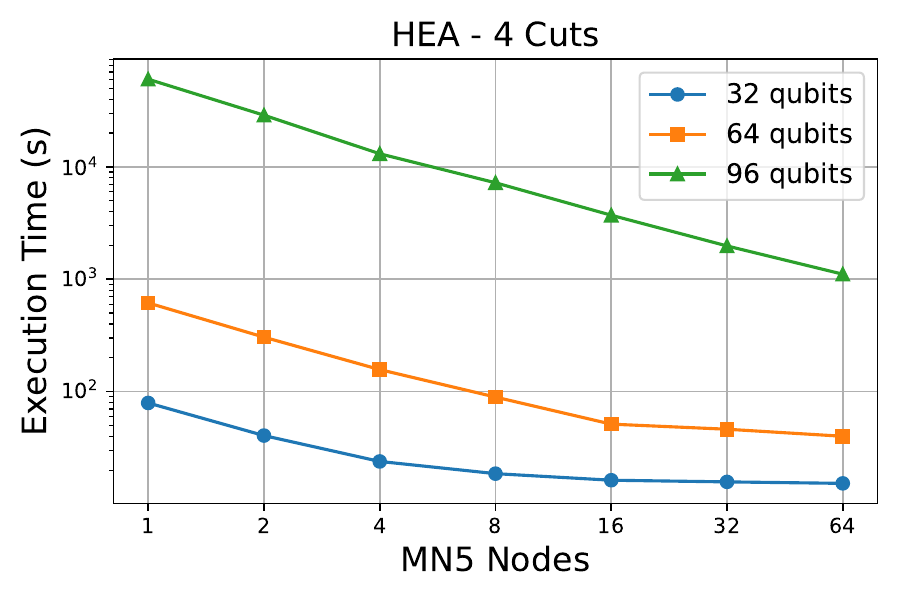}
    \caption{Execution time for HEA circuits with 4 cuts.}
    \label{fig:hea-4cuts}
    
\end{figure}
We first perform the experiments on HEA circuits in MN5-GPP running only on CPUs; Figure~\ref{fig:hea-4cuts} shows the simulation time for reconstructing circuits of 32, 64, and 96 qubits, each cut into 4 fragments of the same number of qubits. The number of qubits per fragment scales with the circuit size—8, 16, and 24 qubits per fragment, respectively. As expected, simulation time increases significantly with the number of qubits. The logarithmic scale in the plot highlights how the growth in simulation time becomes particularly pronounced beyond 64 qubits.  Notably, the scalability across more nodes (from 1 to 64 nodes, thus from 112 to 7,168 cores) is more efficient for larger circuits (e.g., 96 qubits) than for smaller ones (e.g., 32 qubits). This is because larger circuits, when partitioned, benefit substantially more from parallel execution across multiple nodes. The higher number of qubits allows for a more effective distribution of the computational load, whereas smaller circuits may not leverage parallelism, yielding a less substantial decrease in simulation time. 

\begin{figure}[htbp]
    \centering
    \includegraphics[width=0.7\linewidth]{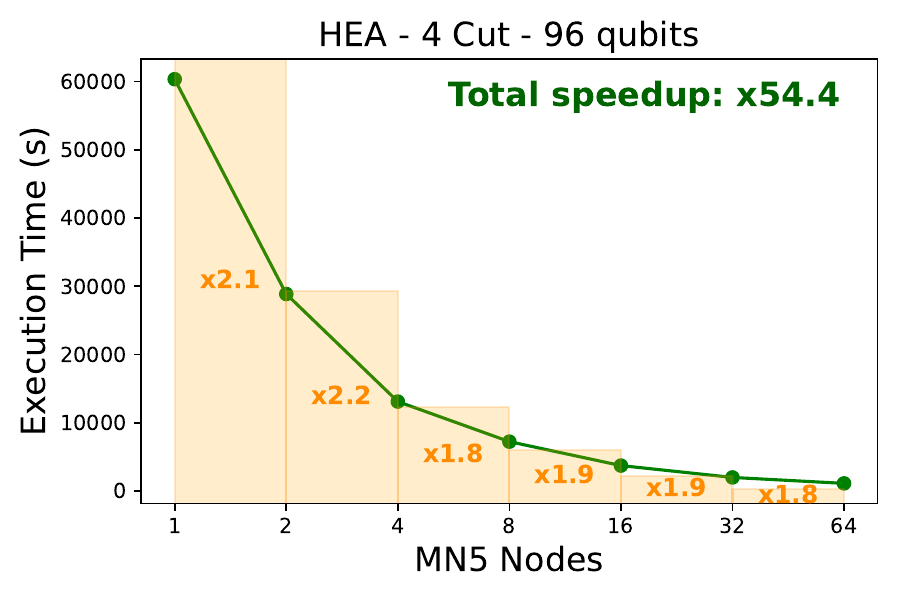}
    \label{fig:enter-label}
    \caption{Speedup for the 96-qubit HEA circuit with 4 cuts.}
    \label{fig:96qubit-speedup}
    
\end{figure}

Figure \ref{fig:96qubit-speedup} presents a detailed view of the 96-qubit case, plotting the simulation time on a linear scale and showing the corresponding speedup achieved as we increase the number of nodes from 1 to 64. The bar chart in the background illustrates the relative speedup between each configuration, culminating in a final speedup of 54.4$\times$ when using 64 nodes—
an excellent figure very close to the ideal (linear) speedup.
Superlinear speedups are observed in the first two transitions: from 1 to 2 nodes and from 2 to 4 nodes. This effect is due to the specific resource distribution in our setup—on each node, the master node uses  12 cores for the PyCOMPSs runtime, leaving 100 cores for task computation. When moving from 1 to 2 nodes, the number of effective worker cores jumps from 100 to 212. 
This explains the sharp initial speedup. As more nodes are added, the speedup continues to grow but at a gradually diminishing rate, approaching the ideal line. This behavior confirms that circuit cutting combined with parallel execution has the potential to scale better, especially when simulating large quantum circuits.

\begin{figure}[htbp]
    \centering
    \includegraphics[width=0.7\linewidth]{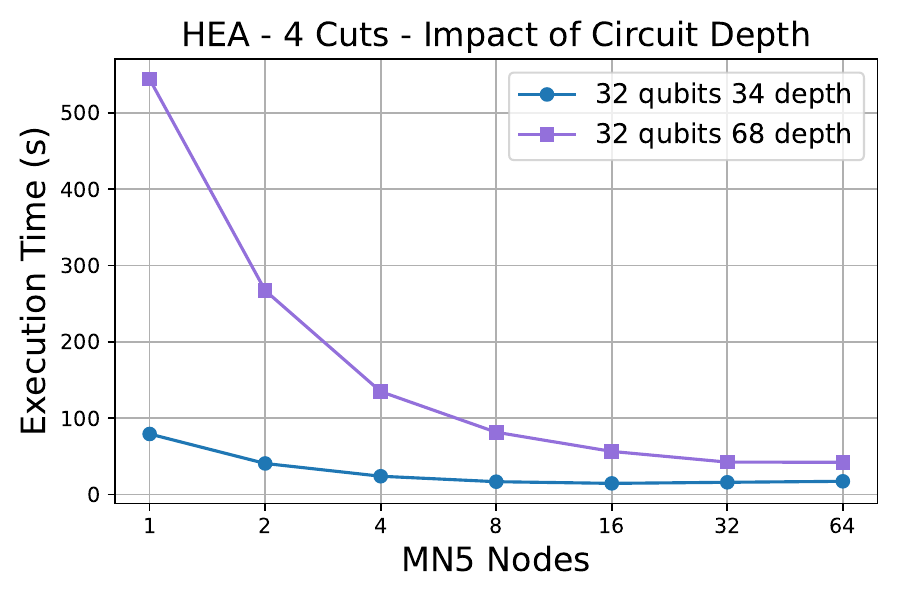}
    \caption{Execution time for HEA circuits varying depth.}
    \label{fig:depth-comparison}
    
\end{figure}

To better understand the effect of circuit depth on simulation performance, we analyze a 32-qubit HEA circuit with two different depths: 34 and 68. In both cases, we apply a total of four cuts. However, due to the doubling of the circuit depth, each individual cut now requires two cuts to isolate subcircuits cleanly. As a result, the circuit with depth 34 is divided into four 8-qubit subcircuits, whereas the deeper circuit (depth 68) is divided into only two 16-qubit subcircuits. This increase in fragment size significantly raises the simulation cost, especially for low node counts, as shown in Figure \ref{fig:depth-comparison}. Nonetheless, the plot also highlights that as we increase the number of nodes, the simulation times for both depths converge, demonstrating that deeper circuits—despite their increased complexity—can still be efficiently simulated with adequate parallel resources. This result shows the strong scalability of the circuit cutting approach, even under more demanding conditions.

\begin{figure}[htbp]
    \centering
    \includegraphics[width=1\linewidth]{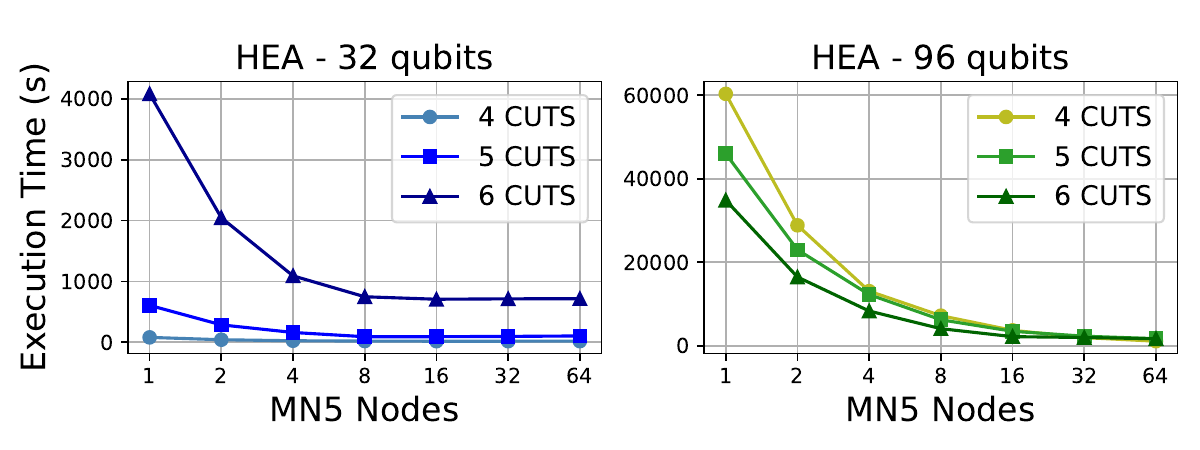}
    \caption{Execution time for HEA circuits varying cuts.}
    \label{fig:cut-comparison}
    
\end{figure}

Figure \ref{fig:cut-comparison} compares simulation times when applying $4$, $5$, and $6$ cuts to circuits of $32$ and $96$ qubits (left and right plots, respectively). In both cases, the number of fragments and their size depend on the number of cuts. For the $32$-qubit circuit, $4$ cuts yield $4$ subcircuits of $8$ qubits; $5$ cuts result in two subcircuits of $7$ qubits and $3$ of $6$ qubits; and $6$ cuts produce two subcircuits of $6$ qubits and $4$ of $5$ qubits.
The left plot shows that increasing the number of cuts in this smaller circuit introduces considerable overhead, leading to longer simulation times—especially in the 6-cut case—and reduced scalability as more nodes are added. This is because the benefit of smaller subcircuits is outweighed by the cost of managing more fragments and performing the reconstruction. On the other hand, the right plot, corresponding to the 96-qubit circuit, shows the opposite trend: the configuration with 6 cuts (producing two subcircuits of $26$ qubits and four of $25$) outperforms the others. Even a $1$-$2$ qubit reduction per fragment significantly improves performance for larger circuits, where simulation cost grows exponentially with qubit count. This highlights a crucial insight: the optimal number of cuts depends heavily on circuit size and properties. While additional cuts may degrade performance in small circuits, they can greatly improve efficiency in large-scale simulations.

\begin{figure}[h]
    \centering
    \includegraphics[width=1\linewidth]{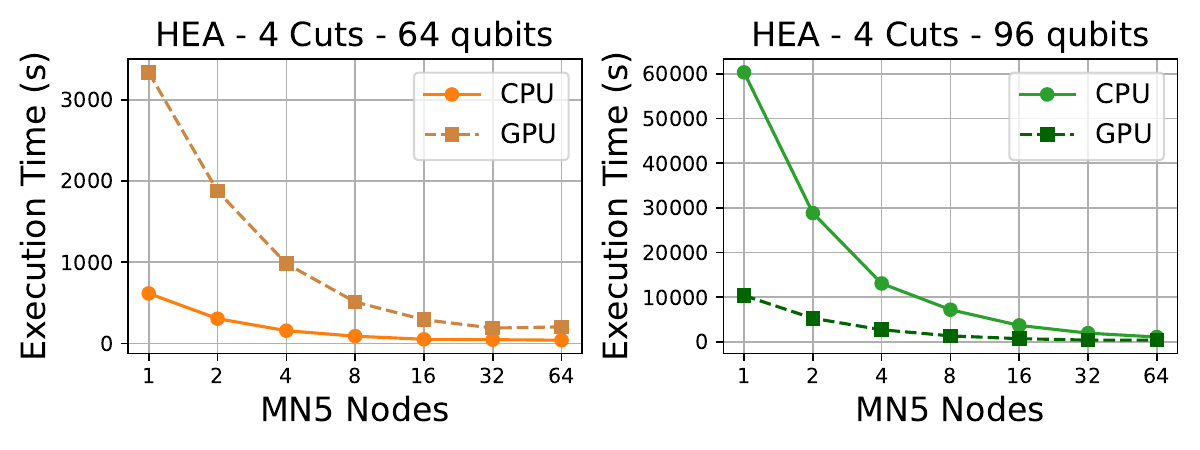}
    \caption{Execution times for $64$-qubit and $96$-qubit circuits with 4 cuts on CPU ($112$ cores) versus GPU (4 gpus).}
    \label{fig:cpu-vs-gpu}
\end{figure}

Figure \ref{fig:cpu-vs-gpu} compares the simulation times of 64-qubit and 96-qubit circuits with 4 cuts, executed on CPU versus GPU, in MN5-GPP and MN5-ACC respectively. In both cases, we run from 1 to 64 nodes, with 112 to 7,168 cores in MN5-GPP and from 4 to 256 GPUs in MN5-ACC. Even though the two type of nodes are not comparable, the goal is to evaluate the behavior of our library in the two architectures. For the 64-qubit circuit on the left, the CPU outperforms the GPU because the smaller circuits are faster to simulate on the CPUs (112 simulations can run concurrently in each node), while the MN5-ACC nodes, with 4 GPUs, face memory transfer overheads and limited parallelization opportunities. This makes the CPU more efficient for smaller circuits. However, for the 96-qubit circuit on the right, the GPU outperforms the CPU as the larger circuit allows the GPU to better utilize its parallel architecture. This demonstrates that GPUs perform better on larger circuits, as they can effectively leverage the parallel processing capabilities of the four available GPUs, significantly reducing the time required to execute a single subcircuit. Both plots show convergence as the number of nodes increases, with the 96-qubit case showing a more pronounced convergence, highlighting the scalability benefits of adding more nodes. 

\subsection{Benchmark 2: Random Circuits}
\begin{figure}[htbp]
    \centering
    \includegraphics[width=1\linewidth]{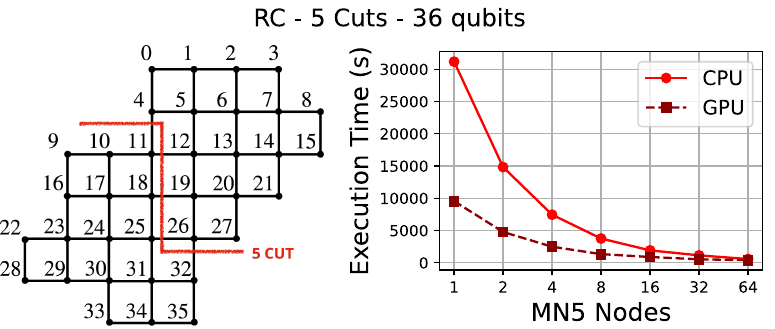}
    \caption{Random Circuit for $36$ qubits. 
    }
    \label{fig:google-36qbits}
\end{figure}

For the second benchmark, we use the Google quantum supremacy circuit~\cite{arute2019quantum}. Our goal is to benchmark the \textit{FindCut} algorithm using a highly connected quantum circuit that also has remarkable applications in demonstrating the computational advantage of quantum computers. In the original quantum circuit, the entangling gates employed are the fermionic simulation gates ($fSim$). Since the focus of this work is to test the optimal cutting scheme and evaluate the computational scaling, we simplify the circuits used by replacing the original $fSim$ gates with CZ gates, whose gate-cutting protocol is simpler. Because we do not use the $fSim$ gate, we refer to the resulting circuits as \textit{Random circuits} (RC). However, the same benchmark could be executed by either finding an optimal gate-cutting protocol for the $fSim$ gate or by decomposing it into CZ and single-qubit gates. The connectivity of the simulated circuit is shown in Figure~\ref{fig:google-36qbits} (left). 

\begin{figure}[htbp]
    \centering
    \includegraphics[width=0.9\linewidth]{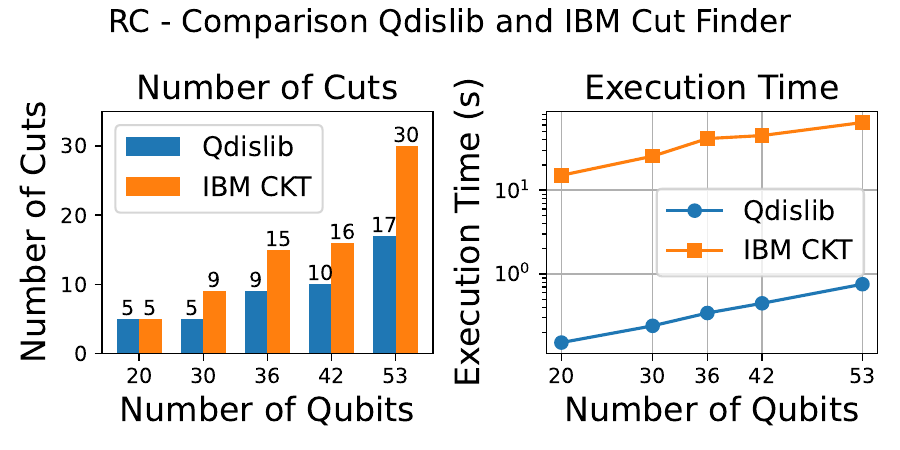}
    \caption{\textit{FindCut} function comparison in \texttt{Qdislib} and  \texttt{IBM CKT} for Random Circuits with a 15-qubit constraint.}
    \label{fig:qdislib-ckt}
\end{figure}

We used this benchmark to first evaluate the performance of the \textit{FindCut}  function. 
Figure~\ref{fig:qdislib-ckt} compares the number of cuts generated  by our \texttt{Qdislib} algorithm against those produced by IBM’s CKT, using the tutorial code from Qiskit’s documentation. We use random circuits with $20$, $30$, $36$, $42$, and $53$ number of qubits, each with a depth of $22$. A constraint is applied to ensure that all resulting subcircuits from the cuts contain less than 15 qubits. As shown in Figure~\ref{fig:qdislib-ckt}, our \textit{FindCut} function is not only faster than IBM’s CKT, but it also finds smaller circuits 
that require less number of cuts. 
These results are comparable to those of other approaches\cite{kan2024scalable}, aligning with the findings in Table II of the same reference, which compares the number of cuts and their efficiency. The faster performance and ability to produce smaller number of cuts underscore the advantages of \texttt{Qdislib} in handling large circuits under specific constraints.

Figure~\ref{fig:google-36qbits} illustrates the execution time of a highly connected 36-qubit quantum circuit with 5 cuts, comparing the performance on CPU versus GPU. In this case, GPUs significantly outperforms CPUs, with a notable reduction in execution time because of the circuits complexity. An interesting aspect of this plot is the convergence behavior of both CPU and GPU simulations. As the number of nodes increases, the execution times for both systems converge very closely. This indicates that, with 64 nodes, both CPU and GPU provide nearly identical performance. This convergence highlights the power of efficient parallelization, suggesting that either hardware can be utilized depending on availability and the specific workload requirements. The GPU's ability to handle larger workloads with better parallelization efficiency is clearly evident, making it the preferred choice for larger-scale and more complex circuits.

\subsection{Hybrid Executions}
\begin{figure}[t]
    \centering
    \includegraphics[width=1\linewidth]{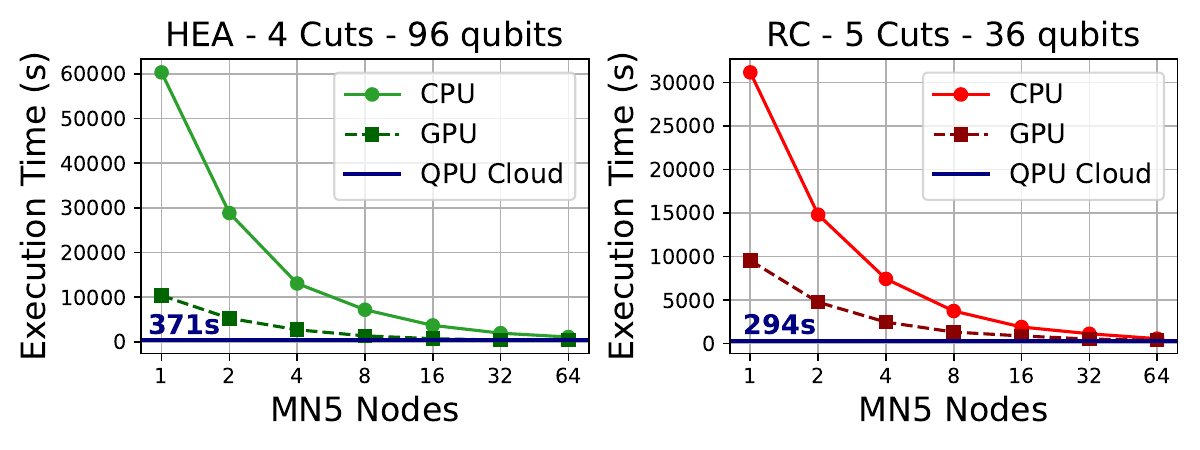}
    \caption{Execution time of circuits with cuts compared to use IBM Cloud QPU with no cuts.}
    \label{fig:hybrid_eval}
\end{figure}

Building on the execution model illustrated in Figure~\ref{fig:hybrid_schema}, we successfully executed a hybrid workflow where subcircuits from a larger quantum circuit are dispatched to different backends (CPU, GPU, and QPU) depending on resource availability and configuration. This section evaluates the performance of such executions. Figure~\ref{fig:hybrid_eval} presents two subplots, already analyzed, comparing CPU and GPU simulation times for 96-qubit and 36-qubit circuits, respectively. Additionally, each subplot includes a third line representing the execution time of the entire circuit (without any cuts) run directly on the IBM Quantum Cloud using the \texttt{Qdislib} library. In \texttt{Qdislib}, this executions with no cuts can be configured with an empty cut set, effectively bypassing the circuit cutting process. Notably, the QPU execution times appear as flat lines at low values (in seconds), intersecting with the CPU and GPU lines where they converge — typically around 64 nodes. This convergence suggests that QPUs offer a clear advantage by enabling execution of larger circuits without the cutting overhead, albeit at the cost of hardware-induced quantum errors. 
Nevertheless, \texttt{Qdislib} demonstrates strong scalability, achieving similar execution times across CPUs and GPUs as the number of nodes increases, narrowing the performance gap with the QPU.

\begin{table}[htbp]
\centering
\footnotesize 
\resizebox{0.95\linewidth}{!}{%
\begin{tabular}{@{}cccccccc@{}}
\toprule
 & \textbf{Qubits} & \textbf{Cuts} & \textbf{CPUs} & \textbf{GPUs} & \textbf{QPU} & \textbf{Cloud QPU} & \textbf{Time} \\
 & & & ncores & ngpus & nqubits & nqubits & (s)\\
\midrule
\textbf{HEA} & 10  & 2 & 112 & -- & -- & -- & 7.1 \\  
\textbf{HEA} & 10  & 2 & --  & 4  & -- & -- & 31.1 \\ 
\textbf{HEA} & 10  & 2 & --  & -- & 5  & -- & 992 \\ 
\textbf{HEA} & 10  & 2 & --  & -- & -- & 5  & 1324 \\ 
\textbf{HEA} & 32  & 3 & 112 & -- & 5  & 5  & 2061 \\ 
\textbf{HEA} & 32  & 3 & 80  & 4  & 5  & 5  & 1597 \\ 
\textbf{HEA} & 64  & 2 & --  & 4  & -- & 39 & 786 \\ 
\textbf{HEA} & 96  & 2 & --  & 4  & -- & 71 & 803 \\ 
\textbf{HEA} & 128 & 2 & --  & 4  & -- & 103 & 826 \\ 
\textbf{RC}  & 36  & 5 & 80  & 4  & -- & -- & 19347 \\
\textbf{RC}  & 36  & 3 & --  & -- & 3  & 33 & 937 \\ 
\textbf{RC}  & 30  & 3 & 112 & -- & -- & 18 & 1318 \\ 
\textbf{RC}  & 30  & 3 & --  & 4  & -- & 18 & 1251 \\
\textbf{RC}  & 30  & 3 & 80  & 4  & 5  & 5  & 1636 \\ 
\textbf{RC}  & 30  & 3 & --  & -- & 5  & 25 & 854 \\ 

\bottomrule
\\
\end{tabular}
}
\caption{Hybrid results}
\label{tab:hybrid}
\end{table}

Table \ref{tab:hybrid} presents several hybrid executions combining different resources—CPUs, GPUs, local QPUs (Ona), and remote QPUs (IBM Cloud)—across several configurations. These results highlight the feasibility and flexibility of hybrid workflows, particularly in how subcircuits are distributed. It is important to note that QPU executions are inherently sequential, as both Ona and IBM Cloud provide access to only one chip per job, meaning that the total execution time is often dominated by the quantum resources. Executions using only Ona tend to be faster than those using only IBM Cloud, underlining the importance of integrating quantum hardware within supercomputing centers to minimize latency and queue time. In these experiments, large subcircuits are typically offloaded to GPUs, smaller ones to Ona (due to its 5-qubit constraint), a limited number to IBM Cloud (due to cost restrictions), and others to CPUs. This approach allows for precise control over subcircuit distribution, showcasing the strength of hybrid executions in resource-aware scheduling. While only a single node of GPU or CPU was used to compensate for the sequential execution on the QPU, the system can be further scaled for larger workloads. Overall, the table demonstrates the adaptability of hybrid computing in efficiently orchestrating complex quantum-classical workloads.

\begin{tcolorbox}[colback=blue!5!white,
colframe=bluebsc,colbacktitle=bluebsc,
title={Hybrid executions highlights:}]
These experiments demonstrate the flexibility of \texttt{Qdislib} to support hybrid executions with HPC systems and quantum devices both on-site and via cloud access. 
A concurrent access is supported thanks to the underlying runtime which also provides good performance figures due to its parallelization support.
\end{tcolorbox}

\section{Related work}
\label{sec:state_of_art}
Circuit cutting appears as a technique to divide larger circuits into smaller ones and reconstructing them post-execution. 
The same circuit cutting technique can be applied to reduce the size of the circuit to be simulated, due to the memory and time limitations when simulating large quantum circuits. This work uses the quantum circuit cutting techniques as presented in \cite{peng2020simulating} and \cite{mitarai2021constructing}. However, there exist other relevant references that present these methods with other terminology or that extend its capabilities to quantum communications. In particular for wire cutting one can also find \cite{tang2021cutqc, 10313822, lowe2023fast, brandhofer2023optimal}, and for gate cutting \cite{ufrecht2023cutting,bechtold2023investigating, fujii2022deep}.

In the context of parallel and distributed computing, \cite{kan2024scalable} proposes FitCut, an approach based on Qiskit to cut circuits and execute them on multi-node quantum systems, which also includes a scheduling algorithm that optimizes resource utilization across nodes.
The article proposes a static algorithm that decides where and when to execute each circuit in each of the quantum systems, while in our case we have been using existing PyCOMPSs scheduling policies to dynamically decide where each circuit is simulated or executed. Future work can propose a scheduling policy in PyCOMPSs for multi-node quantum systems.  

Techniques to perform scalable quantum circuit simulation on multi-node GPU systems have been proposed~\cite{xu2024atlas}. The approach  minimizes  communication costs by partitionining the quantum circuit into a number of stages, each consisting of a (contiguous) subcircuit of the input circuit  simulated on a single GPU. The whole simulation is demonstrated to scale up to 256 GPUs. The parallelization is based on the Flexflow library, implemented on top of Legion. 
Our approach is different to this article with regard two aspects: first, we are proposing the use of traditional circuit cutting to split the circuits while they partition the quantum circuits into stages following a different approach; and second, we can use multiple backends, not only GPUs: CPUs, GPUs, real or cloud-based quantum systems and hybrid backends. 

CutQC~\cite{tang2021cutqc} is another approach to simulate large Quantum Circuits in smaller devices based on circuit cutting techniques. To find the cut locations, it proposes an algorithm based on mixed-integer programming, which may suffer of large execution times when the problem grows.  
In CutQC, the reconstruction phase is executed following a parallel approach, although it is not well defined in the paper and only evaluated up to 16 nodes of low core-count processors.  CutQC was incorporated in the IBM Qiskit circuit cutting addon, but since release 0.9.0 is no longer available. 
The IBM CKT used in this article for comparison purposes is based on the Qiskit addon for circuit cutting~\cite{qiskit-addon-cutting}.

\section{Conclusions}
\label{sec:conclusions}
In this work, we presented \texttt{Qdislib}, a high-performance, distributed quantum library designed for hybrid HPC-quantum executions. We used quantum circuit cutting as the first feature of this library leveraging its applications in quantum algorithms scalability and its parallelizable nature. 
We tested the performance of \texttt{Qdislib} by employing two representative benchmarks - HEA and Random Circuits - for quantum computing applications. We execute these benchmarks in a highly hybrid environment composed by CPUs, GPUs and QPUs,  accessed both locally and through the cloud. We demonstrate efficient scalability when executing on CPU and GPU, as well as the efficiency of an hybrid quantum-classical execution. 

The graph-based representation of quantum circuits used at the core of \texttt{Qdislib}, together with the open-source task-based programming model PyCOMPSs used to orchestrate the executions, makes this tool flexible and hardware-agnostic. 

Quantum computers are still too small and prone to errors to test most real-world applications and thus, circuit cutting is useful both in the short and long term. In this context, \textit{quantum emulation} (quantum simulation in classical devices) has emerged as a near-term solution by simulating quantum applications in a noise-free environment using traditional computation. Even though emulation is limited to a few qubits in general, the quantum advantage frontier can be extended in favor of classical computation if these resources are managed properly. 
\texttt{Qdislib} demonstrates how this task can be made efficient, as shown in the almost ideal speed-up of the executions in CPU and GPU. Moreover, while this work used two standard quantum simulators, \texttt{Qdislib} is modular and can be extended to incorporate other backends. For example, simulators based on Tensor Networks \cite{Ors2014} are highly parallelizable as was shown in \cite{sanchez2021rosnet}, also based in PyCOMPSs, making it natural to merge it with \texttt{Qdislib} in future research.  

Other improvements in \texttt{Qdislib} are possible by exploring reductions in the number of subcircuits needed for each cut. Although the scaling remains exponential with the number of cuts, certain optimization strategies can be adopted, depending on the structure of the circuits \cite{Schmitt2025cuttingcircuits} or the measurement technique applied \cite{lowe2023fast}. Further improvement for the \textit{FindCut} algorithm may include incorporating qubit quality metrics from the quantum hardware or chip topology constraints. Finally, \texttt{Qdislib} can be extended to other quantum computing applications beyond circuit cutting such as variational quantum algorithms or quantum error correction, both of which require an efficient management of quantum and classical computing resources.

In summary, we developed an open-source\footnote{The release of \texttt{Qdislib} as open-source can be found in \newline
\hspace*{1.5em} \url{https://github.com/bsc-wdc/qdislib}}, flexible, hardware-agnostic, and distributed High-Performance quantum library that can execute quantum algorithms in hybrid computational environments. With \texttt{Qdislib}'s benchmarks on real quantum and HPC infrastructures, wendemonstrated full quantum-classical integration and contributed to extending the capabilities of high-performance computing.

\section*{Acknowledgement}
M.T., B.C., A.C.-L. acknowledge funding from the Spanish Ministry for Digital Transformation and of Civil Service of the Spanish Government through the QUANTUM ENIA project call - Quantum Spain, EU through the Recovery, Transformation and Resilience Plan – NextGenerationEU within the framework of the Digital Spain 2026. A.C.-L. acknowledges funding from Grant RYC2022-037769-I funded by MICIU/AEI/10.13039/501100011033 and by “ESF+”.

R.B., J.C. and M.T. acknowledge funding from projects  CEX2021-001148-S, and PID2023-147979NB-C21 from the  MCIN/AEI and MICIU/AEI /10.13039/501100011033 and by FEDER, UE, by the Departament de Recerca i Universitats de la Generalitat de Catalunya, research group MPiEDist (2021 SGR 00412).

Authors acknowledge the use of IBM Quantum Credits for this work. The views expressed are those of the authors, and do not reflect the official policy or position of IBM or the IBM Quantum team.

\bibliographystyle{unsrt}
\bibliography{qdislib}
\end{document}